\documentclass[10pt,journal,compsoc]{IEEEtran}
\newif\ifpeerreview

\peerreviewtrue

\usepackage[nocompress]{cite}
\usepackage{url}
\usepackage{amsmath,amssymb,graphicx}

\usepackage{lipsum} 
\usepackage[center]{caption}

\usepackage[switch]{lineno}

\title{Removing Blocking Artifacts in Video Streams Using Event Cameras}

\author{Henry~Chopp*,
        Srutarshi~Banerjee,
        Zihao~Wang,
        Oliver~Cossairt,
        and~Aggelos~K.~Katsaggelos
\IEEEcompsocitemizethanks{\IEEEcompsocthanksitem M. Shell is with the Department
of Electrical and Computer Engineering, Georgia Institute of Technology, Atlanta,
GA, 30332.\protect\\
E-mail: see http://www.michaelshell.org/contact.html
\IEEEcompsocthanksitem J. Doe is with Anonymous University.}
}

\begin{document}

\IEEEtitleabstractindextext{%
\begin{abstract}
In this paper, we propose EveRestNet, a convolutional neural network designed to remove blocking artifacts in video streams using events from neuromorphic sensors. We first degrade the video frame using a quadtree structure to produce the blocking artifacts to simulate transmitting a video under a heavily constrained bandwidth. Events from the neuromorphic sensor are also simulated, but are transmitted in full. Using the distorted frames and the event stream, EveRestNet is able to improve the image quality.
\end{abstract}

}

\author{Henry~Chopp*,
        Srutarshi~Banerjee,
        Oliver~Cossairt,
        and~Aggelos~K.~Katsaggelos
\IEEEcompsocitemizethanks{\IEEEcompsocthanksitem H. Chopp, S. Banerjee, and A.K. Katsaggelos are with the Department of Electrical and Computer Engineering at Northwestern University.\protect\\
*E-mail: HenryChopp2017@u.northwestern.edu
\IEEEcompsocthanksitem O. Cossairt is with the Department of Computer Science at Northwestern University.}}
{}
\maketitle

\IEEEraisesectionheading{
  \section{Introduction}\label{sec:introduction}
}
%
%
%
%
\IEEEPARstart{N}{euromorphic} sensors recently have been gaining in popularity.
Unlike traditional color/intensity cameras, these sensors record changes in the log-intensity at each pixel.
If a change exceeds the preset threshold, a so-called event is reported as either an increase (say, +1) or a decrease (-1), depending on whether the intensity increases or reduces respectively. The binary nature of these event cameras is of course a drawback in comparison with other conventional cameras that capture a much wider dynamic range. Each event is represented as a tuple of $(x, y, t, p)$ where $(x,y)$ represent the 2 dimensional coordinate of the event fired, while $t$ and $p$ represent the timestamp and the polarity of the event fired at that location.
However, there are a few key engineering trade-offs that these novel devices provide: (i) asynchronous firing of events, (ii) event latency on the order of $10\mu s$, (iii) low power consumption on the order of $10 mW$, and, (iv) redundant capture of static scenes.
These benefits open up new paths in solving various vision problems. Event cameras have brought new solutions to many classical as well as novel problems in computer vision and robotics, including high frame-rate video reconstruction \cite{ed-vfs,scheerlinck2020fast,shedligeri2019photorealistic}, with HDR \cite{eventHDR2019,rebecq2019high} and high resolution \cite{wang2020eventsr,choi2019learning,gef}, and 3D reconstruction of human motion \cite{xu2019eventcap} and scenes \cite{rebecq2018emvs,kim2016real}, as well as odometry \cite{censi2014low, vidal2018ultimate} and tracking \cite{zhu2017tracking, lagorce2014asynchronous}.

For example, switching from a traditional camera to an event-based camera would give longer life to battery-operated robots.
Simultaneous Localization and Mapping (SLAM) applications have been tested with event cameras in robotics.
Multimodal applications are pursued as well: high speed video frame interpolation.

In traditional video compression standards such as  \cite{VVC1},\cite{VVC2}, \cite{hevc}, \cite{h_264}, the video is compressed using a quadtree (QT) based compression. At very high distortions corresponding to low bit rates, the quality of the video frames suffer from blocking artifacts. This results in edges having a block-like appearance. These blocking artifacts results in not only poor visual quality of the objects in the scene, but also reduces efficient intelligent analytics in the scene such as object detection or tracking in the scene using deep learning based approaches. Typically, the performance of the neural network based methods work better with high quality frames compared to frames with low bit rates and poor quality. In order to address this issue, one of the possible techniques would be the removal of the blocking artifacts in the frames. This often can be solved by image restoration or quality improvement of frames using deep learning approaches such as Generative Adversarial Network (GAN) approaches. On the other hand, the asynchronous events occurring due to relative motion between camera and scene can be used intelligently in order to remove these blocking artifacts from the captured video frames. To the best of authors' knowledge, there has been no prior work in literature addressing this problem using events. We are therefore, the first to attempt such a problem of removing the blocking artifacts in a video captured at low bit rate using events. 

In this paper, we propose a deep learning based approach in order to perform a restoration of low bit rate videos. The deep learning model comprises of a neural network with residual blocks. The model takes in events occurring between time $t-1$ and $t$. The model takes the previous restored frame at time $t-1$ as input as well. The model generates restored frame at $t$. The main contribution of the paper are as follows:
\begin{itemize}
    \item Development of a deep learning framework to remove blocking artifacts using events.
    \item Image restoration performance at different distortion levels.
    \item Performance comparison with other image restoration techniques.
\end{itemize}

In this work, we focus on restoring image frames with blocking artifacts due to video compression at low bit rates. The framework can be extended onto other restoring applications such as de-blurring, super-resolution and others. One of the limitations of this approach is the poor reconstruction quality when there is lack of events in the scene, due to small motion or lack of texture in the scene.

\section{Related Work}
While we believe we are the first to try image restoration in the context of removing blocking artifacts, there have been many publications that address aspects of the problem we face. Wang et al. \cite{ed-vfs} for example used non-distorted video feeds along with events to interpolate video frames. Baura et al. \cite{baura2016} and Rebecq et al. \cite{e2vid} were early adopters of using learning methods to reconstruct intensity images from only events.

A plethora of papers address the removal of blocking artifacts typically from block-based compression schemes such as JPEG. Within the last few years, deep learning has allowed for the problem to resurface. Dong et al. \cite{dong2015compr} was one of the first to use deep learning to solve this very issue. Lin et al. \cite{lin2019blocking} more recently also used deep learning techniques that showed improvements in the PSNR of the restored images.

These methods, however, do not quite address the fusion problem that we face: (i) we are looking to \textit{restore} degraded intensity images using events, not generate intensity images from events, and (ii) the blocking artifacts we are addressing arise from quadtree (QT) compression, which is inherently different than JPEG compression in terms of the size of the blocks and the values that are used to represent the blocks. JPEG compression stores information from the discrete Fourier transform (DFT) for each block, while QT compression stores a singular intensity value that fills the entirety of the size-varying blocks.

\subsection{Quadtree Compression Scheme}
The compression scheme we use to generate blocking artifacts is a QT-based approach proposed by Banerjee et al. \cite{reimagine}. A host-chip architecture is used to analyze large video feeds. The chip, which consists of a conventional intensity camera, a neuromorphic camera, and low processing power, must compress each frame before sending it to the host computer over heavily constrained bandwidth. QTs are used as the method of compression, and due to the low bandwidth, must transmit lossy frames.

\begin{figure}[ht]
    \centering
    \includegraphics[width=0.45\linewidth]{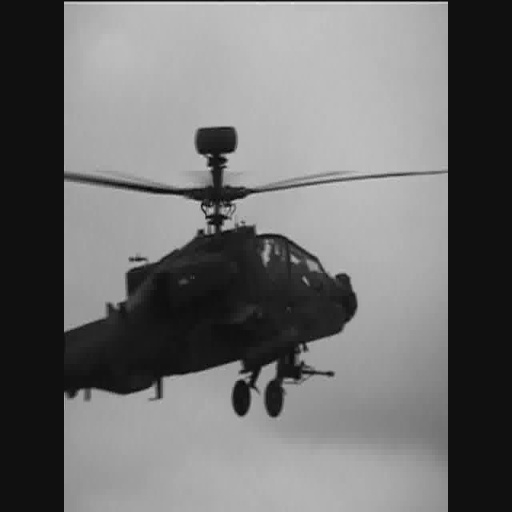}
    \includegraphics[width=0.45\linewidth]{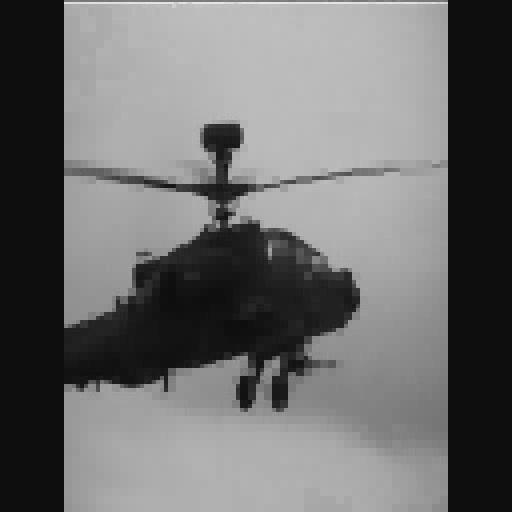}
    \caption{(Left) Undistorted frame (Right) Distorted frame}
    \label{fig:sample_frame}
\end{figure}

The host computer is assumed to possess as much computational power as needed in order to perform object tracking on the degraded feeds. Based partially on the locations of the objects of interest, the host communicates back the areas where the chip should more finely divide the QT. The video feeds on the host follow this pattern: objects of interest typically have more QT blocks (i.e. better sampling) than that of the background and objects not of interest. A sample frame from the ImageNet Vid dataset \cite{ILSVRC15} is shown, undistorted and distorted, in Fig. \ref{fig:sample_frame}. Notice how the binning has produced blocking artifacts particularly along the high spatial frequency components: the helicopter blades have noticeably been degraded, and looking closely one can see the jagged edges of the helicopter body.

In order to perform object tracking using deep neural networks, it's well-known that testing or run-time data drawn from a similar distribution as the training data will perform better than data drawn from a different distribution. If we have a preprocessing step that can transform the data back to its original form, then subsequent analysis is generally more accurate. In this work, we want to improve the similarity (e.g. PSNR) between the original undistorted image and the distorted image.

\section{Proposed Method}
Our goal is as follows: we would like to restore video frames degraded with blocking artifacts by means of events from neuromorphic sensors. Assume we have (i) a degraded intensity video feed $\hat{\mathcal{F}}_t = \{\hat{f}_0, \hat{f}_1, \ldots, \hat{f}_t\}$ composed of degraded frames $\hat{f}_i$ up to time $t$, and (ii) low-latency events $\mathcal{E}_t$ that contain the set of all events up to time $t$. Original undistorted frames $\mathcal{F}_t = \{f_0, f_1, \ldots, f_t\}$ are compressed using the system proposed in \cite{reimagine}. Denote an individual event at time $t$ by $e_t = (x, y, p)$ where $(x,y)$ is the location of the event and $p \in \{-1, +1\}$ is the event polarity. At any given pixel, an event is fired if the difference in log-intensity goes above ($p = +1$) or below ($p = -1$) a predetermined threshold. Using these data, we can greatly improve the quality of the images using deep learning.

\subsection{EveRestNet}
\begin{figure*}[ht]
    \centering
    \includegraphics[width=\textwidth]{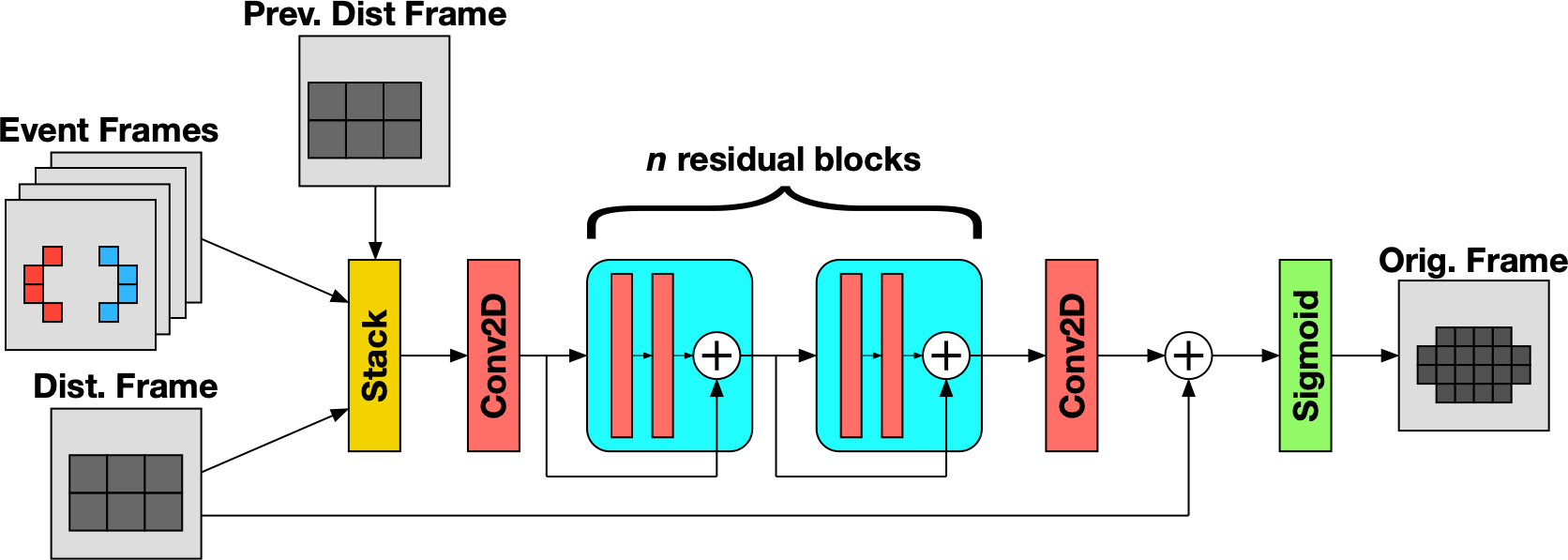}
    \caption{Architecture of EveRestNet}
    \label{fig:everestnet}
\end{figure*}

Here we propose EveRestNet, a convolutional neural network (CNN) that uses (eve)nts to (rest)ore blocking artifacts that appear in intensity-based video feeds. The architecture as seen in Fig. \ref{fig:everestnet} is derived from the successes of ResNet \cite{resnet} where we attempt to learn the residual detail lost in the original distorted frame. Additionally, since EveRestNet is fully convolutional, it can accept any video resolution.

\subsubsection{Data Input}
We use a subset of the data available at time $t$. Since we want to only focus on restoring $\hat{f}_t$, it makes sense to also utilize the previous distorted frame $\hat{f}_{t-1}$ along with all the events that were fired between times $t-1$ and $t$, i.e. $E_t = \mathcal{E}_t - \mathcal{E}_{t-1}$. Events are fired asynchronously, meaning that the timestamps cannot be discretized without losing information on when they were fired. However, EveRestNet is convolutional, so all dimensions need to be discretized in order to be valid inputs into the neural network. We found that temporally binning events in $E_t$ using four bins worked well. These four event frames we will denote as $E_{t,i}$ where $i \in I = \{1,2,3,4\}$ is an event frame identifier within $E_t$.

The final EveRestNet architecture requires six inputs: $X_t = \{\hat{f}_t,\ \hat{f}_{t-1},\ E_{t,i}\ \forall i \in I\}$. These inputs are all of the same spatial size, and are all concatenated channel-wise as a data volume that is passed into the network.

\subsubsection{Loss Function}
The areas in the degraded frames with the lowest signal-to-noise ratio is typically along the high-frequency components where the edges are not as smooth as the ground truth images. When trying to restore the edges, we want to pay particular attention to those edges since it is where the large errors tend to lie.

\begin{figure}[ht]
    \centering
    \includegraphics[width=0.24\linewidth]{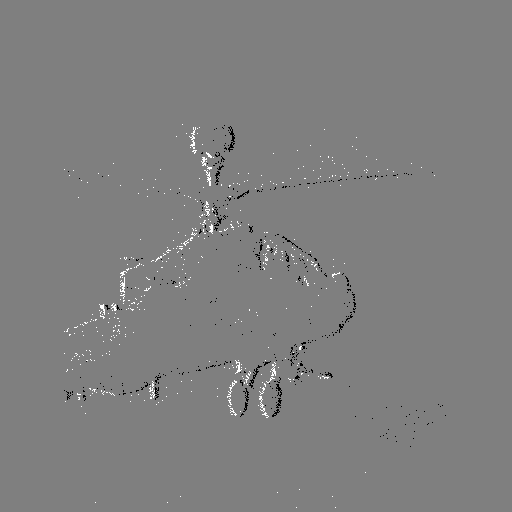}
    \includegraphics[width=0.24\linewidth]{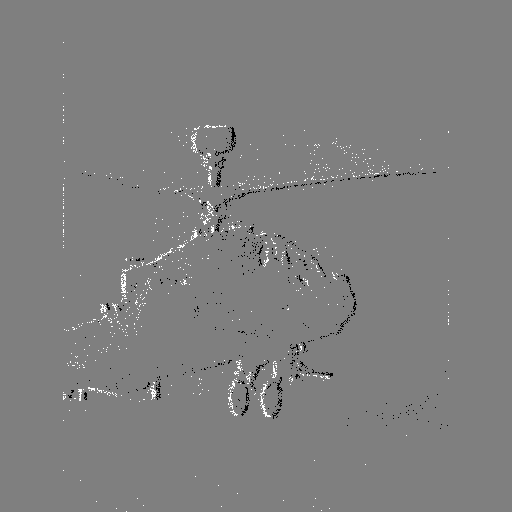}
    \includegraphics[width=0.24\linewidth]{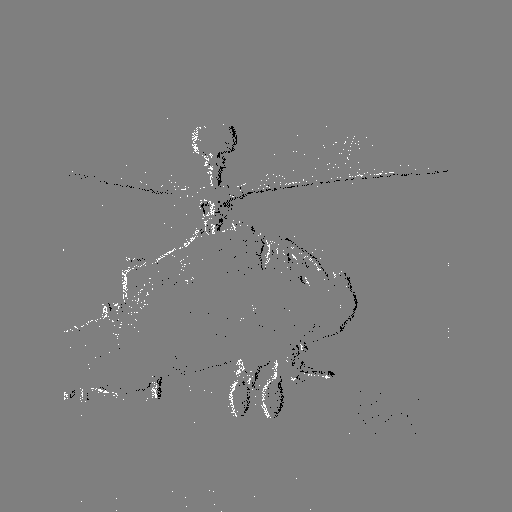}
    \includegraphics[width=0.24\linewidth]{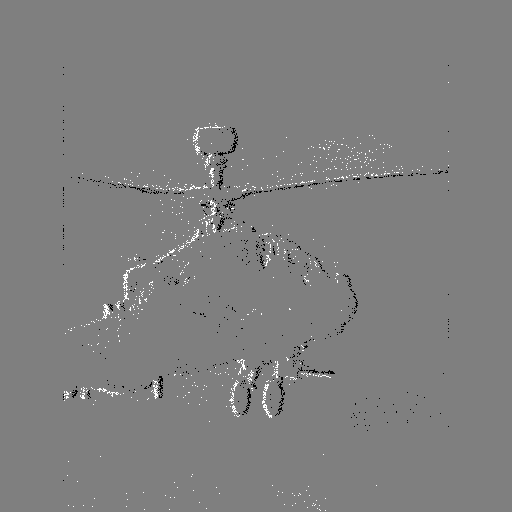}
    \caption{Four binned event frames corresponding to Fig. \ref{fig:sample_frame}}
    \label{fig:ev_frames}
\end{figure}

Events can be thought of as a threshold of temporal gradients in intensity at each pixel. As the objects move in the video feeds, edges and other textures high in spatial frequency tend to trigger the firing of events. This can be seen in the four event frames of Fig. \ref{fig:ev_frames}. These event frames correspond to the events that occur between Fig. \ref{fig:sample_frame} and its previous frame. Events with $p = -1$ are represented as black pixels, and events with $p = +1$ are white pixels. Gray denotes no events in the pixel's location.

The event frames can hold information on the smoothness of the object boundaries, and should be used to our advantage. When training, we use a loss function that emphasizes the importance of events in restoring the distorted frames.

Let $r_t = EveRestNet(X_t)$ be the restored output of EveRestNet. Our loss function is defined as the sum of two losses: a weighted fidelity term, and a weighted total variation term. The total loss $\mathcal{L}$ is

\begin{equation}\label{eq:total_loss}
    \mathcal{L} = \mathcal{L}_{fid} + \mathcal{L}_{TV}
\end{equation}
where
\begin{align}
    \mathcal{L}_{fid} = \left\Vert\left(1+\lambda_{fid}\bar{E}_t\right)*\left(f_t - r_t\right)\right\Vert_2^2 \label{eq:l_fid}\\
    \mathcal{L}_{TV} = \left\Vert\lambda_{TV}\left(4 - \bar{E}_t\right) * \nabla r_t\right\Vert_2^2 \label{eq:l_tv}
\end{align}
and where $\bar{E}_t = \sum_{i = 1}^4\left|E_{t,i}\right|$ represents the number of times an event fired at each pixel, and $\nabla r_t$ represents the spatial gradient of the restored image. Weighting parameters $\lambda_{fid}$ and $\lambda_{TV}$ are chosen experimentally.

The fidelity term $\mathcal{L}_{fid}$ is given extra weight at pixels where there were more events, which typically indicate the presence of edges. Again, the edges are the areas where most of the restoration needs to happen. The total variation term $\mathcal{L}_{TV}$ gives lesser weight to areas without events. We do not want to suppress the edges where the events occurred; however, areas without events tend to be low in texture and can appropriately be smoothed.

\begin{figure*}[ht]
    \centering
    \begin{tabular}{c c c c c}
     & $E_{t,4}$ & $\hat{f}_t$ & $f_t$ & $r_t$ \\
     \rotatebox{90}{\footnotesize ILSVRC2015\_train\_01054000} &
     \includegraphics[width=0.22\linewidth]{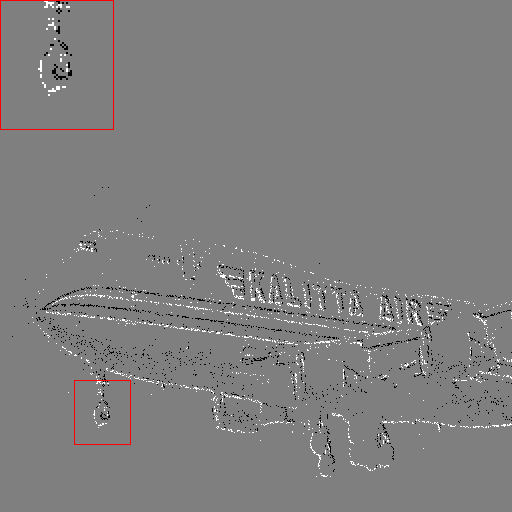} &
     \includegraphics[width=0.22\linewidth]{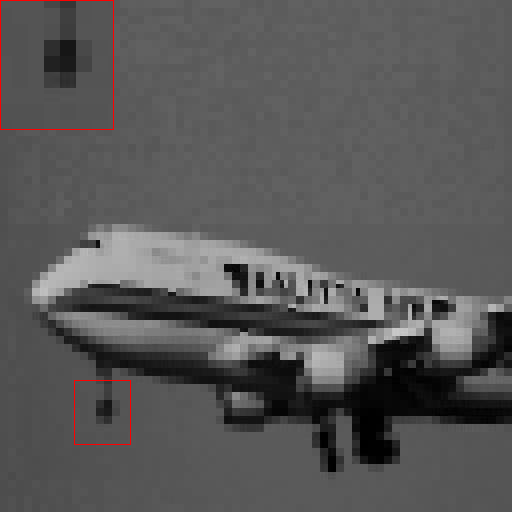} &
     \includegraphics[width=0.22\linewidth]{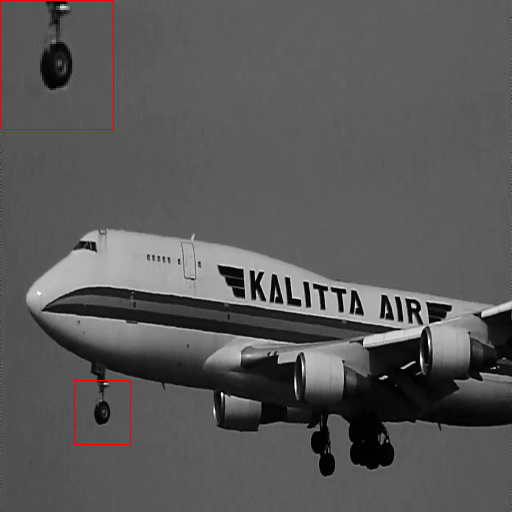} &
     \includegraphics[width=0.22\linewidth]{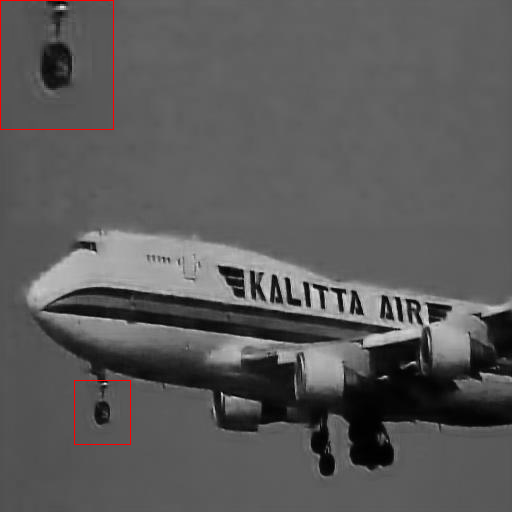} \\
     & & PSNR: 20.70 (23.23) & & PSNR: 30.11 (30.87) \\
     & & SSIM: 0.7893 (0.8119) & & SSIM: 0.9339 (0.9182) \\
     \\
     
     \rotatebox{90}{\footnotesize ILSVRC2015\_train\_00515000} &
     \includegraphics[width=0.22\linewidth]{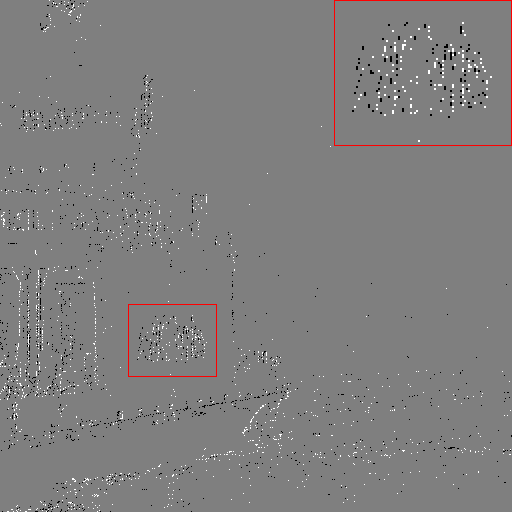} &
     \includegraphics[width=0.22\linewidth]{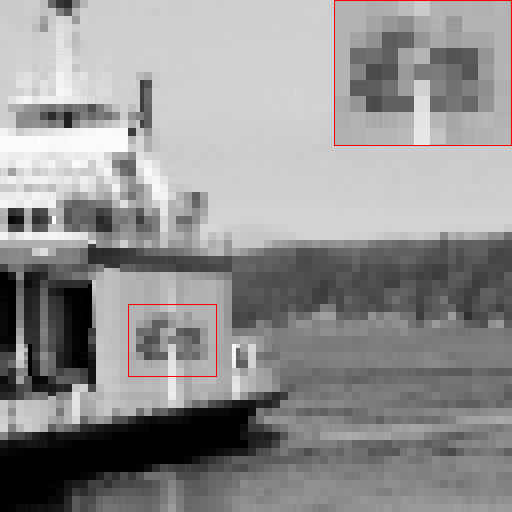} &
     \includegraphics[width=0.22\linewidth]{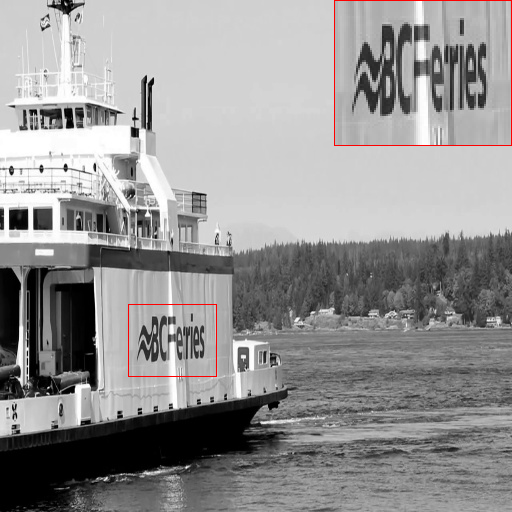} &
     \includegraphics[width=0.22\linewidth]{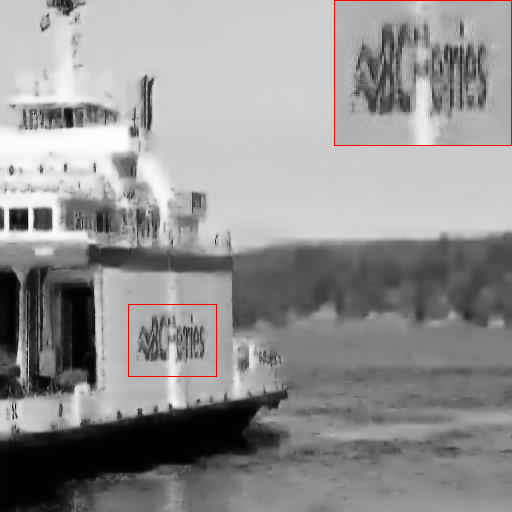} \\
     & & PSNR: 20.43 (16.60) & & PSNR: 23.33 (20.44) \\
     & & SSIM: 0.5995 (0.4267) & & SSIM: 0.6877 (0.6980) \\
     \\
     
     \rotatebox{90}{\footnotesize ILSVRC2015\_train\_00047000} &
     \includegraphics[width=0.22\linewidth]{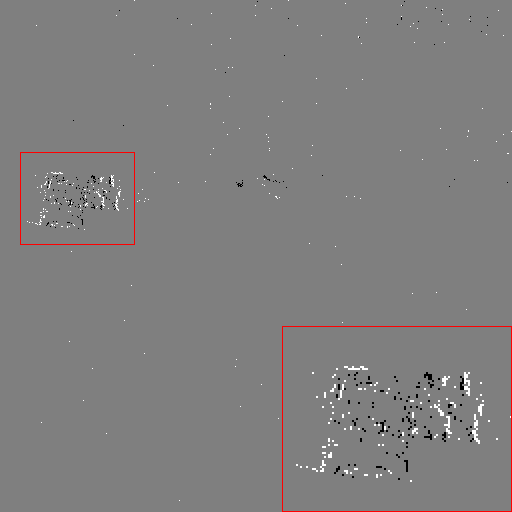} &
     \includegraphics[width=0.22\linewidth]{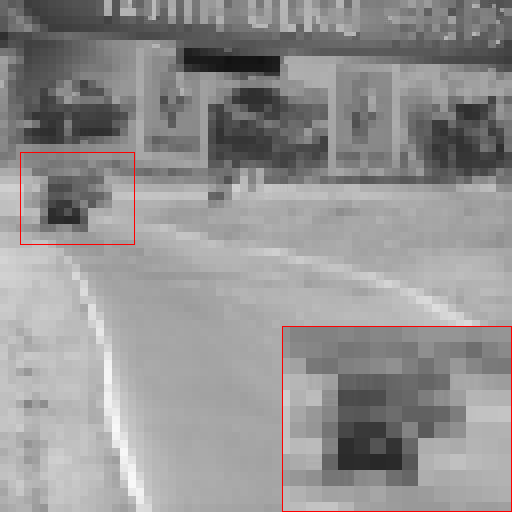} &
     \includegraphics[width=0.22\linewidth]{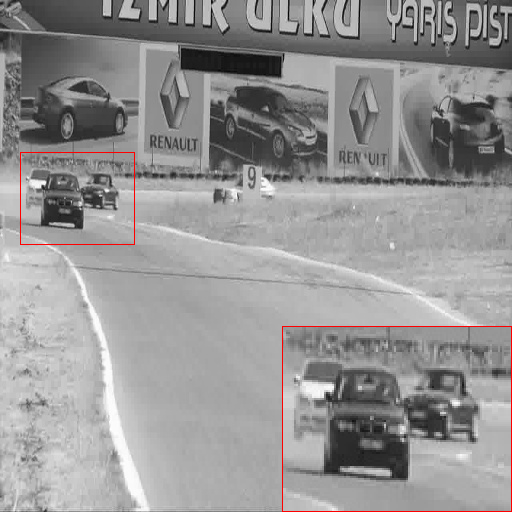} &
     \includegraphics[width=0.22\linewidth]{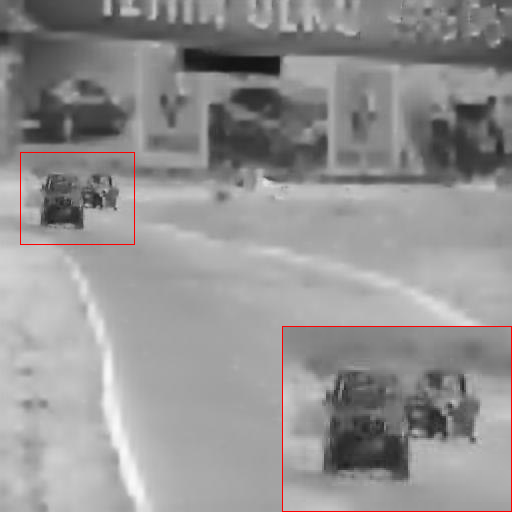} \\
     & & PSNR: 23.49 (18.62) & & PSNR: 24.38 (21.93) \\
     & & SSIM: 0.6680 (0.4716) & & SSIM: 0.7188 (0.6269) \\
    \end{tabular}
    \caption{Sample results of EveRestNet. PSNR and SSIM for the frame and highlighted region (in parentheses) is shown.}
    \label{fig:results}
\end{figure*}

\section{Experimental Results}
Our results show promising improvements in restoring the frames of the degraded video sequences. Below we discuss the training and evaluation of EveRestNet.

\subsection{Training EveRestNet}
We trained EveRestNet using a subset of data from the ImageNet Vid dataset \cite{ILSVRC15} with video feeds of airplanes, cars, and watercraft. This dataset does not provide corresponding event data, so we turned to ESIM \cite{Rebecq18corl} to generate simulated events for the video feeds. The simulated events were then temporally binned into four frames per each pair of consecutive image frames.

The degraded frames were generated using the process described in \cite{reimagine} where the objects of interest are the three classes chosen to train EveRestNet with (which is also the same three classes Banerjee et al. used). These objects in general have a finer binning sizes than that of the background, but still experiences blocking artifacts much like what is seen in Fig. \ref{fig:sample_frame}. Different degrees of degradation were generated by varying the bandwidth constraints. This is to ensure the network can be robust to changes in the amount of distortion. 

The final version of EveRestNet uses 4 residual blocks each with size $3\times3\times32$ convolution kernels. The padding and step size of 1 allows for any sized input. Batch normalization is applied after each convolution within a residual block. Leaky ReLU activation functions with slope $0.2$ appear after the first convolution of EveRestNet as well as after the first convolution of each residual block. Gradient $\nabla r_t$ is generated using the $3\times3$ Sobel filters in the $x$- and $y$-directions. We set $\lambda_{fid} = 0.5$ and $\lambda_{TV} = 0.05$. EveRestNet was optimized using Adam with a learning rate of $5\times10^{-4}$, $\beta_1 = 0.9$, and $\beta_2 = 0.999$.

\subsection{Results}
After training, we tested EveRestNet on 1500 frames of varying degradations. We compare the peak signal-to-noise ratio (PSNR) and the structural similarity index measure (SSIM) of the degraded frame $\hat{f}_t$ and the frame generated by EveRestNet, $r_t$. On average, we achieved a higher PSNR: the average PSNR of all $\hat{f}_t$ is $24.91$, while EveRestNet achieved an average PSNR of $29.05$. For SSIM, $\hat{f}_t$ averaged $0.7583$ while EveRestNet averaged $0.8479$.

Some figures are shown in Fig. \ref{fig:results}. To save space and show detail, only the event frame closest temporally to time $t$, $E_{t,4}$, is shown. The previous degraded frame $\hat{f}_{t-1}$ is also omitted. While there are portions of the results that are enhanced to show the fine detail in the results, we invite the reader to zoom in to see other details captured by EveRestNet that would otherwise be lost using the degraded frame alone.

The sequences shown in Fig. \ref{fig:results} have stationary fixed camera angles (or effectively stationary in the case of the plane with no discernible background). This is why we see smoothing of edges in the background where no events are fired, whereas in the moving objects themselves we see that the blocking artifacts are effectively removed. In the airplane, the lettering becomes readable, and finer details such as the door and passenger windows become visible. The boat too has a logo that is recovered using EveRestNet. The car sequence in the highlighted portion of Fig. \ref{fig:results} is able to detail two of the three cars present. Perceptually, the frames are much more informative using EveRestNet than the degraded frames alone. Both quantitative metrics demonstrate significant improvement in the image quality as well.

\section{Conclusion}
We faced the issue of how to improve the image quality of video frames impacted by blocking artifacts. The frames are degraded from a bandwidth-constrained system that uses QT-based methods to compress the image. The system, if equipped with a neuromorphic sensor, can provide additional information in the form of an event stream to combat the issue. EveRestNet, our proposed solution, is a CNN that utilizes the distorted frames along with the events to restore images closer to the undistorted ground truth video frames. Our preliminary results of EveRestNet show just how much we can improve the video quality with these neuromorphic sensors.

\ifpeerreview \else
\section*{Acknowledgments}
The authors would like to thank...
\fi

\bibliographystyle{IEEEtran}
\bibliography{references}

\ifpeerreview \else


\begin{IEEEbiography}{Michael Shell}
Biography text here.
\end{IEEEbiography}


\begin{IEEEbiographynophoto}{John Doe}
Biography text here.
\end{IEEEbiographynophoto}


\fi

\end{document}